\documentclass[english,aps,prl,twocolumn]{revtex4-1}
\usepackage[T1]{fontenc}
\usepackage[latin9]{inputenc}
\usepackage{amsmath}
\usepackage{graphicx}
\usepackage{esint}

\makeatletter
\@ifundefined{textcolor}{}
{%
 \definecolor{BLACK}{gray}{0}
 \definecolor{WHITE}{gray}{1}
 \definecolor{RED}{rgb}{1,0,0}
 \definecolor{GREEN}{rgb}{0,1,0}
 \definecolor{BLUE}{rgb}{0,0,1}
 \definecolor{CYAN}{cmyk}{1,0,0,0}
 \definecolor{MAGENTA}{cmyk}{0,1,0,0}
 \definecolor{YELLOW}{cmyk}{0,0,1,0}
}

\makeatother

\usepackage{babel}
\begin{document}

\title{Many-particle dephasing after a quench}

\author{Thomas Kiendl$^{1,2}$}

\author{Florian Marquardt$^{2,3}$}

\affiliation{$^{1}$Dahlem Center for Complex Quantum Systems and Institut für
Theoretische Physik, Freie Universität Berlin, 14195, Berlin, Germany}

\affiliation{$^{2}$Institute for Theoretical Physics, Universität Erlangen-Nürnberg,
Staudtstr. 7, 91058 Erlangen, Germany}

\affiliation{$^{3}$Max Planck Institute for the Science of Light, Günther-Scharowsky-Straße
1/Bau 24, D-91058 Erlangen, Germany}
\begin{abstract}
After a quench in a quantum many-body system, expectation values tend
to relax towards long-time averages. However, in any finite-size system,
temporal fluctuations remain. It is crucial to study the suppression
of these fluctuations with system size. The particularly important
case of non-integrable models has been addressed so far only by numerics
and conjectures based on analytical bounds. In this work, we are able
to derive analytical predictions for the temporal fluctuations in
a non-integrable model (the transverse Ising chain with extra terms).
Our results are based on identifying a dynamical regime of 'many-particle
dephasing', where quasiparticles do not yet relax but fluctuations
are nonetheless suppressed exponentially by weak integrability breaking. 
\end{abstract}
\maketitle
\emph{Introduction} \textendash{} The relaxation dynamics of quantum
many-body systems has come under renewed scrutiny in the past years,
due to its relevance for the foundations of thermodynamics and the
availability of isolated systems, like cold atoms. The simplest case
considers the evolution after a sudden quench of parameters \cite{polkovnikov2011colloquium}.
Typically, one then analyzes local physical observables (like particle
density, magnetization, currents), and asks about the time-evolution
of expectation values. The most basic question concerns the long-time
\emph{averages} after the quench: are they correctly described by
a thermal state at some effective temperature related to the initial
energy after the quench? \cite{PhysRevLett.98.050405,Rigol:2008aa,PhysRevA.43.2046,PhysRevE.50.888,PhysRevLett.80.1373,PhysRevLett.98.210405,PhysRevLett.100.030602,PhysRevLett.98.180601,PhysRevLett.97.156403,PhysRevLett.100.100601,PhysRevE.85.060101,2014arXiv1403.6481T,PhysRevE.82.031130,eisert2015quantum,PhysRevLett.113.050601,1994cond.mat.10046S}.
On the next, more refined level of analysis we can study the time-dependent
\emph{fluctuations} of expectation values around their temporal average.
For any finite system, these persist even at infinite time. In principle,
these represent a kind of long-term memory, since they are reproducible
(the same for each repetition of the quench) and depend both on the
exact time of the quench and on details of the initial state. 

A crucial question for the foundations of statistical physics is:
are these fluctuations suppressed in the thermodynamic limit $N\rightarrow\infty$,
and if yes, how fast? This is also relevant for experiments in equilibration,
like analog quantum simulations carried out in finite ('mesoscopic')
lattices.

These fluctuations around the time-average are commonly characterized
by $\sigma_{A}^{2}=\overline{[\langle\hat{A}(t)\rangle-\langle\hat{{A}}\rangle_{\text{eq}}]^{2}}$
\cite{PhysRevLett.101.190403,1367-2630-13-5-053009,PhysRevE.87.012106,PhysRevE.50.888}.
The overbar denotes a time-average and $\langle\hat{{A}\rangle}_{\text{eq}}=\overline{\langle\hat{A}(t)\rangle}$.
Note that this is different from the quantum fluctuations ${\rm Var}\hat{A}(t)=\langle[\hat{A}(t)-\langle\hat{A}(t)\rangle]^{2}\rangle$,
which are usually much larger and would be present even in a perfect
thermal equilibrium state (where $\sigma_{A}^{2}$ vanishes).

The finite-size scaling of persistent temporal fluctuations after
a quench has been approached so far from several angles: (i) in the
context of the Eigenstate Thermalization Hypothesis, justifying the
neglect of off-diagonal contributions to expectation values \cite{srednicki1994,Srednicki1999,rigol2008,PhysRevA.80.053607},
(ii) based on the former, general mathematical bounds supplemented
by physical arguments for generic interacting, non-integrable systems
\cite{PhysRevLett.101.190403,1367-2630-13-5-053009,1402-4896-86-5-058512,1367-2630-14-1-013063};
(iii) calculations for simple integrable systems (which have, however,
special properties that strongly differ from the generic case) \cite{PhysRevE.87.012106,PhysRevE.89.022101,PhysRevLett.106.140405,PhysRevA.86.053615,PhysRevE.81.061134};
(iv) numerics \cite{PhysRevE.88.032913}. 

Here, we will provide exact analytical results for the suppression
of fluctuations in a \emph{non-integrable} system, confirming the
hypothesized exponential decay with system size. Our analysis rests
on identifying a general dynamical regime which we term '\emph{many-particle
dephasing}', relevant for weak integrability breaking. The advantage
over having purely numerical results will be that we can provide a
complete description of how the result depends on the quench, the
initial state, and parameters. The advantage vs. analytical bounds
is that the bounds are not guaranteed to be close to the true results. 

\emph{Integrable transverse Ising Model} \textendash{} We start from
the well-known integrable quantum Ising chain. We review briefly its
properties and its quench dynamics, as they will be important for
our analytical solution of the non-integrable evolution later on.
The quantum (transverse) Ising chain is an exactly solvable model
for quantum phase transitions \cite{Lieb1961,Pfeuty1970,calabrese2006,sachdev2007phaseTransitions,PhysRevA.2.1075,PhysRevLett.85.3233}:
\begin{equation}
\hat{H}_{0}=\frac{\Omega}{2}\sum_{j=1}^{N}\hat{\sigma}_{z,j}-J\sum_{j=1}^{N}\hat{\sigma}_{x,j}\hat{\sigma}_{x,j+1}
\end{equation}
Here $\hat{\sigma}_{x,j}$ and $\hat{\sigma}_{z,j}$ are spin-1/2
operators acting on site $j$. We will assume periodic boundary conditions,
with $\hat{\sigma}_{x,N+1}=\hat{\sigma}_{x,1}$. For $J<\Omega/2$,
the model is paramagnetic (where $\left\langle \hat{\sigma}_{z,j}\right\rangle <0$),
while at $J=\Omega/2$ there is a quantum phase transition into a
ferromagnetic phase, with spins aligning either in the $+x$ or $-x$
direction. The model can be solved exactly by mapping to free fermions,
via $\hat{\sigma}_{+,j}=\hat{c}_{j}^{\dagger}\exp(i\pi\sum_{l=1}^{j-1}\hat{c}_{l}^{\dagger}\hat{c}_{l})$.
This results in a quadratic fermionic Hamiltonian that does not conserve
particle number and can be solved by Bogoliubov transformation in
$k$-space:
\begin{equation}
\hat{H_{0}}=\sum_{k}(\Omega-2J\cos(k))\hat{c}_{k}^{\dagger}\hat{c}_{k}-Ji\sin(k)(\hat{c}_{k}^{\dagger}\hat{c}_{-k}^{\dagger}-\hat{c}_{-k}\hat{c}_{k})\label{eq:k-spaceHamiltonian}
\end{equation}
For definiteness we will assume $N$ even. The quantization of wavenumbers
is slightly changed from the textbook case (due to an extra sign that
enters when coupling site $N$ to site $1$), with $k=\frac{2\pi}{N}(l+\frac{1}{2})$,
where $l$ is an integer and $k$ ranges over the Brillouin zone $[-\pi,\pi[$.
The Hamiltonian decomposes into independent sectors $(k,-k)$. 

For this as well as other integrable models it has been found that
the temporal variance of many single-particle observables scales like
$1/N$ \cite{PhysRevE.87.012106,PhysRevLett.106.140405}. However,
there are important exceptions where there is no such suppression
with $N$ \cite{PhysRevA.86.053615,PhysRevLett.113.050601}.

In a quench of the coupling strength $J$ out of the pre-quench ground
state, during the evolution we will have $(k,-k)$ either occupied
by two particles or unoccupied. This can be viewed as an artificial
spin 1/2 system. We take $\hat{S}_{zk}=-1$ to correspond to $\left|0_{-k},0_{k}\right\rangle $
and $\hat{S}_{zk}=+1$ representing $\left|1_{-k},1_{k}\right\rangle =\hat{S}_{k}^{+}\left|0_{-k},0_{k}\right\rangle $,
with $\hat{S}_{k}^{+}\equiv\hat{c}_{k}^{\dagger}\hat{c}_{-k}^{\dagger}$.
In that notation, the Hamiltonian (\ref{eq:k-spaceHamiltonian}) becomes
a set of decoupled effective spin-1/2 systems:
\begin{equation}
\hat{H}_{0}=\frac{1}{2}\sum_{k>0}\Omega_{k}\vec{b}_{k}\hat{\vec{S}}_{k}\label{eq:TransverseIsingHamiltonianSpinLanguage}
\end{equation}
We introduced the two-particle excitation frequencies $\Omega_{k}=\sqrt{(2\Omega-4J\cos(k))^{2}+(4J\sin(k))^{2}}$,
and the field direction $\vec{b}_{k}=(0,4J\sin(k),2\Omega-4J\cos(k))^{T}/\Omega_{k}$.
The ground and excited state, $\left|\pm_{k}\right\rangle $, have
energies $\pm\Omega_{k}/2$, and $\left\langle \pm_{k}\left|\hat{\vec{S}}_{k}\right|\pm_{k}\right\rangle =\pm\vec{b}_{k}$.
In this picture, a quench corresponds to a sudden change of $\Omega_{k}$
and $\vec{b}_{k}$, such that for each $k$ the Bloch vector starts
to precess around the new field direction: $d\left\langle \hat{\vec{S}}_{k}\right\rangle /dt=\Omega_{k}\vec{b}_{k}\times\left\langle \hat{\vec{S}}_{k}\right\rangle $. 

An example for an observable is the projector for the spin pointing
along $+z$ at some site $j$: $\hat{A}=\hat{\sigma}_{+,j}\hat{\sigma}_{-,j}$.
Because of translational invariance, $\langle\hat{A}(t)\rangle$
is independent of $j$. One finds $\langle\hat{A}(t)\rangle=N^{-1}\sum_{k>0}(\langle\hat{S}_{zk}\rangle+1)$,resulting
in an expression of the form $\langle\hat{A}(t)\rangle=N^{-1}\sum_{k>0}\left(A_{0k}+A_{ck}\cos(\Omega_{k}t)+A_{sk}\sin(\Omega_{k}t)\right)$.
 At sufficiently long times, all the oscillatory terms dephase, producing
seemingly random time-dependent fluctuations. This process can be
termed ``\emph{single-particle} dephasing'', since it results from
the superposition of different oscillation frequencies whose number
scales \emph{linearly} with system size.

Thus, the temporal variance ends up being $\sigma_{A}^{2}=N^{-2}\sum_{k>0}(A_{ck}^{2}+A_{sk}^{2})/2$.
In the limit of large $N$, this becomes 
\begin{equation}
\sigma_{A}^{2}=N^{-1}\int_{0}^{\pi}\frac{dk}{2\pi}(A_{ck}^{2}+A_{sk}^{2})/2\,,
\end{equation}
i.e. $\sigma_{A}^{2}\sim N^{-1}$, confirming the result of \cite{PhysRevE.87.012106}.

\emph{Quench in the non-integrable model} \textendash{} The general
physical expectation for non-integrable systems is that the long-time
steady state after a quench has fluctuations that are exponentially
suppressed in particle number (system size) $N$, in contrast to the
power-law suppression in the integrable case displayed above. This
was made explicit first in \cite{PhysRevLett.101.190403}. There,
an upper bound was derived, $\sigma_{A}^{2}\leq(a_{{\rm max}}-a_{{\rm min}})^{2}\cdot{\rm IPR}$.
Here, $a_{{\rm max}}$ and $a_{{\rm min}}$ are the maximum and minimum
eigenvalues of $\hat{A}$. ${\rm IPR}=\sum_{n}\left|\left\langle \Phi_{n}\left|\Psi(0)\right.\right\rangle \right|^{4}$
denotes the inverse participation ratio, which decreases if the initial
state $\left|\Psi(0)\right\rangle $ spreads over more energy eigenstates
$\left|\Phi_{n}\right\rangle $. It was then argued on general physical
grounds that the ${\rm IPR}$ usually decreases exponentially with
system size. However, an argument of this kind does not reveal how
fast the decay is for any concrete system or quench scenario, or whether
the upper bound displays the correct parameter dependence at all,
since it will not be tight in general. More recently, it was reported
that numerical simulations for a variety of models and quench scenarios
indeed reveal an exponential suppression with system size, for the
finite-size systems that could be addressed \cite{PhysRevE.88.032913}. 

\begin{figure}
\includegraphics[width=1\columnwidth]{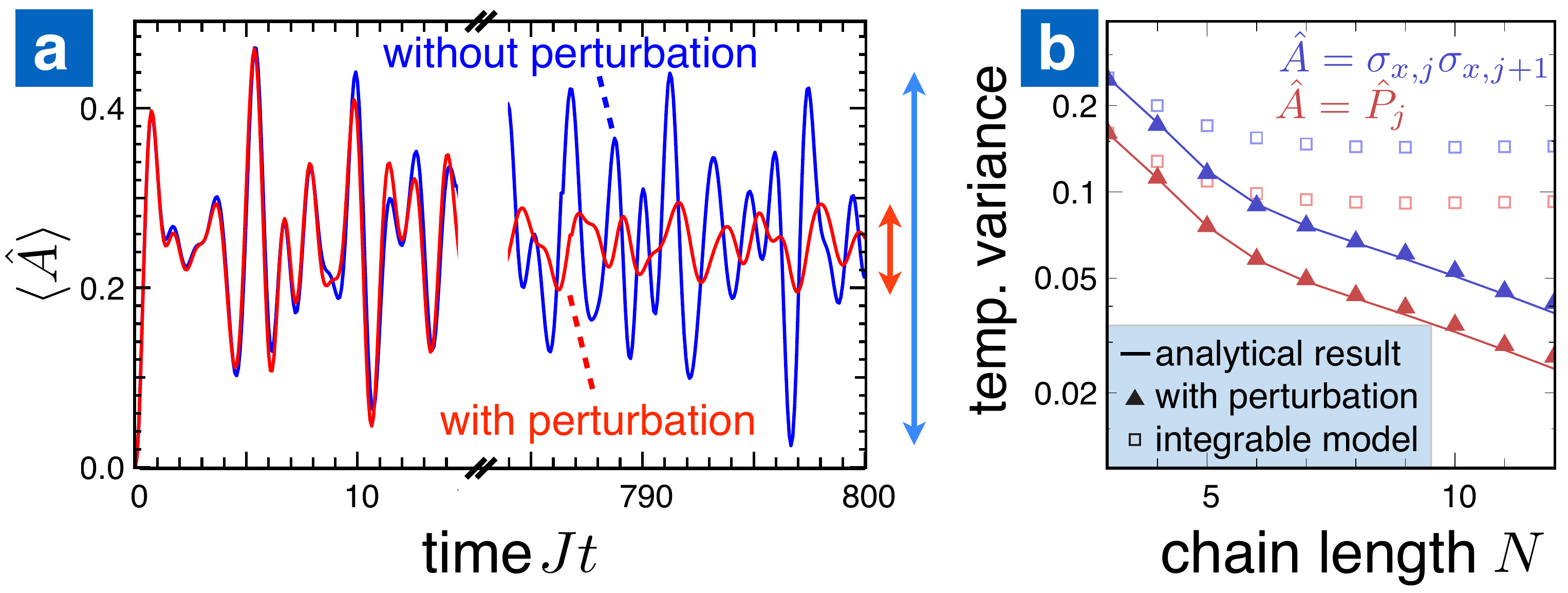}

\protect\caption{\label{fig:Raw dynamics}Many-particle dephasing in a chain with $N=12$.
The quench jumps from $J_{\text{pre}}/\Omega=0$ to $J/\Omega=0.8$. Without
perturbation (blue) the fluctuations at early and late times are similar.
A weak NNN coupling of strength $J_{NNN}/J=0.01$ leads to a significant
additional relaxation for $t\rightarrow\infty$. (b) The temporal
variance $N\sigma_{A}^{2}$ for two different observables shows an
exponential decay in $N$. {[}Analytical result from Eq.~(\ref{eq:ManyBodyDephasing for weak non-integrable models}).{]}}
\end{figure}

Our goal here is to go beyond bounds and numerics, and to find an
analytical expression for a non-integrable case. We break the integrability
of the quantum Ising model by adding next-nearest-neighbor (NNN) coupling
$\hat{H}_{{\rm NNN}}=-J_{{\rm NNN}}\sum_{j}\hat{\sigma}_{x,j}\hat{\sigma}_{x,j+2}$.
In the fermion representation, this gives rise to two-particle interactions.
Other choices for integrability breaking are possible, which we will
address later. A direct numerical simulation (Fig.~(\ref{fig:Raw dynamics}))
indeed reveals a stronger suppression of fluctuations, that seems
to be consistent with an exponential decay in $N$.

We now come to an important question: What is the physical origin
of this strong suppression? Initially, one might suspect 'true thermalization',
in the sense of inelastic scattering of quasiparticles leading to
a redistribution of quasiparticle populations. This process could
then be described using a kinetic equation, and the final state would
be thermal. However, the simulation shows that this is not the case,
the quasiparticle distribution remains practically unchanged. There
is further numerical evidence that we are not witnessing thermalization:
The fluctuations decay to their steady-state long-time limit during
a time-scale $\tau^{*}$ that scales linearly in the inverse perturbation:
$\tau^{*}\sim\left|J_{{\rm NNN}}\right|^{-1}$. This is in contrast
to the behaviour expected from a kinetic equation, where the relaxation
rate would be set by $J_{NNN}^{2}$.

\begin{figure}
\includegraphics[width=1\columnwidth]{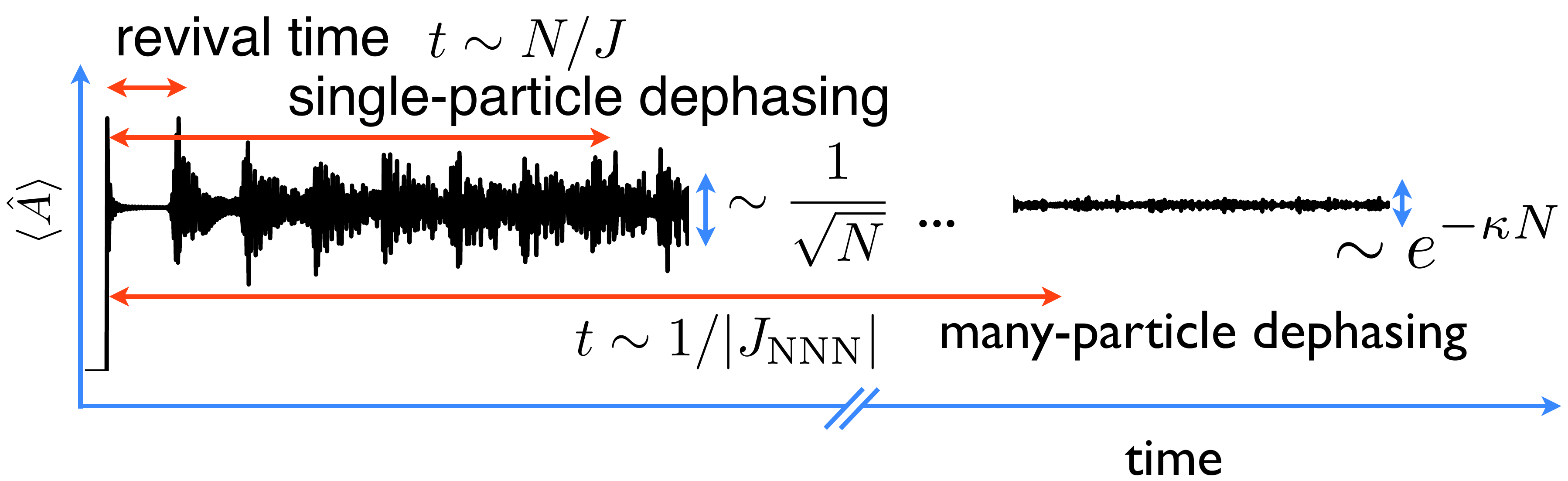}

\protect\caption{\label{fig:Overview}Schematic overview of different dynamical regimes
after a quench in a finite-size system that is weakly perturbed away
from an integrable, effectively non-interacting model (displayed for
a local observable in the perturbed transverse Ising model, but valid
more generally). First, revivals occur. Second, a transient steady-state
with fluctuations $\sigma_{A}\sim1/\sqrt{N}$ is observed, as predicted
for the integrable case. Finally, after a time that scales as the
inverse of the integrability-breaking perturbation, the final steady-state
is reached. There, fluctuations are reduced exponentially in the system
size, due to many-particle dephasing.}
\end{figure}

\emph{Many-particle dephasing} \textendash{} Instead, we have identified
a mechanism that could be termed 'many-particle dephasing'. First,
we note that, for weak interactions, the many-body energy eigenstates
still coincide to a very good approximation with those of the integrable
model. This explains the absence of thermalization in the occupations
of quasiparticles. At the same time, however, the energies are changed.
This lifts the exponentially large degeneracies of the integrable
model and gives rise to dephasing. The number of frequencies involved
is now exponentially large in $N$, which is the reason we term the
resulting dynamics ``many-particle dephasing''. The generic situation,
including the different timescales, is shown schematically in Fig.~\ref{fig:Overview}.
We note that interacting systems mappable to noninteracting ones (the
present case) and purely non-interacting systems have to be distinguished.
Only in the former the complete one-particle density matrix relaxes
\cite{PhysRevLett.113.050601}.

It can be shown easily (e.g. \cite{PhysRevLett.101.190403}) that
fluctuations in the long-time limit obey $\sigma_{A}^{2}=\sum_{\Delta\neq0}\left|\sum_{\Delta_{\alpha}=\Delta}A_{\alpha}\right|^{2}$.
Here $\alpha=(f,i)$ denotes a transition between two energy eigenstates
$i$ and $f$ where $\Delta_{\alpha}=E_{f}-E_{i}$ is the transition
energy, and $A_{\alpha}=\Psi_{f}^{*}A_{fi}\Psi_{i}$ combines the
transition matrix element of the observable with the amplitudes $\Psi_{l}=\left\langle \Phi_{l}\left|\Psi(0)\right.\right\rangle $
of the initial state with respect to the post-quench energy eigenbasis
$\Phi_{l}$. 

We now consider an arbitrary transition $i\rightarrow f$ that is
induced by $\hat{A}$. Suppose the observable just affects a single
quasiparticle at a time, or (in our case) it affects only a single
$(k,-k)$ pair of states. All other quasiparticles (or $k$-pairs)
are merely spectators. Such a structure is typical for single-particle
observables. It is at this point that the weak integrability-breaking
interactions impose a crucial difference. For the integrable (effectively
non-interacting) case, there is an exponentially large number of other
transitions that have the same transition energy. These are obtained
by picking all possible configurations of the remaining 'spectator'
degrees of freedom (which are identical in the initial and final state).
In contrast, for the non-integrable (weakly interacting) case, there
is a correction to the transition energies which lifts this massive
degeneracy. For the present model, the transition energy correction
$\delta\Delta_{fi}\sim J_{{\rm NNN}}$ turns out to be a sum over
contributions that depend on pairs of occupation numbers, $n_{k}$
and $n_{k'}$ (see Suppl. Material). Given a change in one of the
occupation numbers, the correction thus depends on the configuration
of all the 'spectator' degrees of freedom. Therefore, barring any
(rare) accidental degeneracies, the initial degeneracy is completely
lifted. That statement is confirmed by direct numerical inspection
of $\delta\Delta_{fi}$. 

Assuming that all the transition energies $\Delta_{\alpha}$ have
become non-degenerate, we find $\sigma_{A}^{2}=\sum_{f\neq i}\left|\Psi_{f}^{*}A_{fi}\Psi_{i}\right|^{2}$.
In general, it would still be an impossible task to evaluate this
expression analytically. At this stage, however, the important observation
is that the $\Delta_{\alpha}$ do not enter any more, even though
their modification by the weak interaction was crucial to lift the
degeneracies. Our strategy will be to evaluate this expression for
the matrix elements calculated with respect to the unperturbed integrable
many-particle eigenfunctions. In this way, we will arrive at analytical
insights into the suppression of fluctuations for the \emph{non-integrable}
model! The requirement for this to work is that the perturbation $J_{{\rm NNN}}$
is still weak, such that the eigenfunctions have not been changed
appreciably. Later we will check the results against numerics.

Each energy eigenstate of the integrable transverse Ising model can
be written as a product state: $\left|\Phi_{n}\right\rangle =\Pi_{k>0}\left|\varphi(n,k)\right\rangle $.
Each configuration $n$ is described by $N/2$ bits $\varphi(n,k)\in\left\{ -1,+1\right\} $,
where $-1$ denotes the ground state $\left|-_{k}\right\rangle $
and $+1$ the excited state $\left|+_{k}\right\rangle $ in the $(k,-k)$
sector.  The observable we focused on in the numerical example was
$\hat{A}=(\hat{\sigma}_{z,j=0}+1)/2$, which, in fermionic language,
is equal to $\hat{A}=N^{-1}\sum_{k,k'}\hat{c}_{k}^{\dagger}\hat{c}_{k'}\exp[-i(k-k')x]$.
For this observable, we find:
\begin{equation}
\langle\Phi_{m}|\hat{A}|\Phi_{n}\rangle=\frac{2}{N}\sum_{k>0}\left\langle \varphi(m,k)\left|\hat{S}_{k}^{+}\hat{S}_{k}^{-}\right|\varphi(n,k)\right\rangle I_{k}\,,\label{eq:GeneralOverlapFormula}
\end{equation}
where $I_{k}\equiv\Pi_{k'\neq k}\delta_{\varphi(m,k'),\varphi(n,k')}$
enforces the initial and final configurations of 'spectators' $k'\neq k$
to match.

In evaluating the general formula for $\sigma_{A}^{2}$, we have to
sum over all possible many-particle transitions $i\rightarrow f$.
However, the Kronecker delta in Eq.~(\ref{eq:GeneralOverlapFormula})
enforces the configurations $\varphi(i,k')$ and $\varphi(f,k')$
to be equal except at $k'=k$. We still have to sum over exponentially
many configurations, though that can be handled by regrouping terms
and a bit of combinatorics (see Suppl. Material). In doing this, we
exploit the fact that the initial state can be written as a product
state over the different $k$-sectors (since it is an eigenstate of
the pre-quench Hamiltonian).

\emph{Analytical results} \textendash{} The final analytical result
for the long-term, steady-state fluctuations in the weakly non-integrable
model (small $J_{{\rm NNN}}\neq0$) is:
\begin{equation}
\sigma_{A}^{2}=\frac{\mathcal{C}}{N}\cdot\exp\left(-2\kappa N\right)\label{eq:ManyBodyDephasing for weak non-integrable models}
\end{equation}
Here the exponential is equal to the inverse participation ratio ${\rm IPR}$.
We find explicitly $2\kappa=-\frac{1}{N}\sum_{k>0}{\rm ln}\,{\rm IPR(k)}$,
where ${\rm IPR(k)=}\left|\left\langle +_{k}\left|\Psi_{k}\right.\right\rangle \right|^{4}+\left|\left\langle -_{k}\left|\Psi_{k}\right.\right\rangle \right|^{4}$
is the IPR for the initial state $\left|\Psi_{k}\right\rangle $ in
sector $k$. In the limit of large $N$, $\kappa$ becomes $N$-independent:
$2\kappa\rightarrow\int_{0}^{\pi}\frac{dk}{2\pi}{\rm ln}{\rm IPR}(k)$.
Thus, we obtain analytical access to the exponential suppression of
fluctuations. 

The prefactor in $\sigma_{A}^{2}$ contains a further $1/N$ suppression,
and a constant $\mathcal{C}$, which can be given explicitly as well:
\begin{equation}
\mathcal{C}=\frac{8}{N}\sum_{k>0}w(k)\approx8\int_{0}^{\pi}\frac{dk}{2\pi}w(k)\,,
\end{equation}
where
\begin{equation}
w(k)={\rm IPR}^{-1}(k)\cdot\left|\left\langle +_{k}\left|\hat{S}_{k}^{+}\hat{S}_{k}^{-}\right|-_{k}\right\rangle \right|^{2}\cdot P_{k}(1-P_{k})\,,\label{eq:w(k) equation}
\end{equation}
and $P_{k}=\left|\left\langle +_{k}\left|\Psi_{k}\right.\right\rangle \right|^{2}$.

\begin{figure}
\includegraphics[width=1\columnwidth]{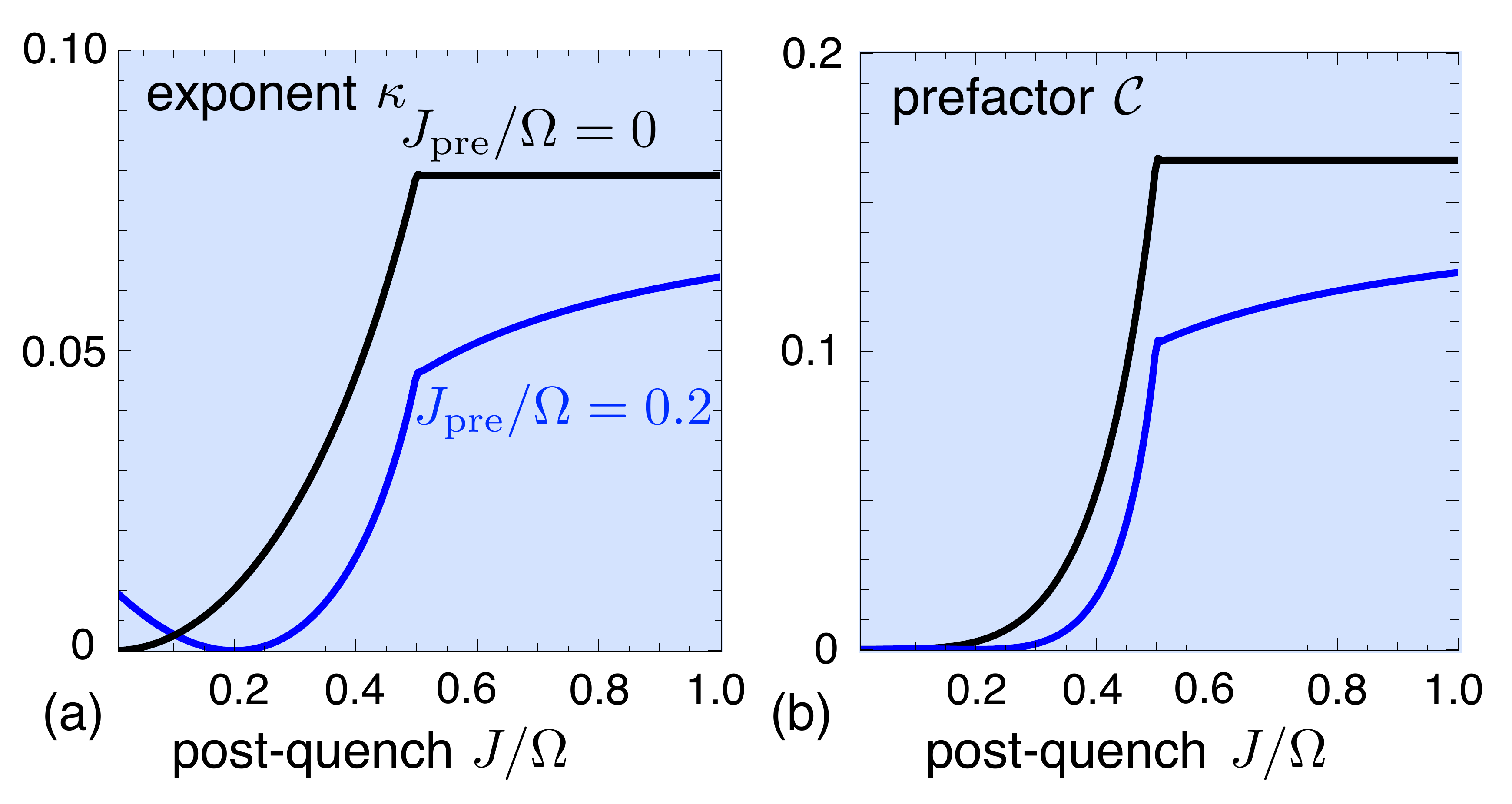}

\protect\caption{\label{fig:DecayConstantAndPrefactor}Analytical predictions, depending
on the post-quench parameter $J$. (a) Decay constant $\kappa$ for
the decay of fluctuations with system size, and (b) prefactor $\mathcal{C}$
(see main text, in the formal limit $N\rightarrow\infty$). The quench
assumed here jumps from a coupling $J_{{\rm pre}}$ to $J$.}
\end{figure}

\begin{figure}
\includegraphics[width=1\columnwidth]{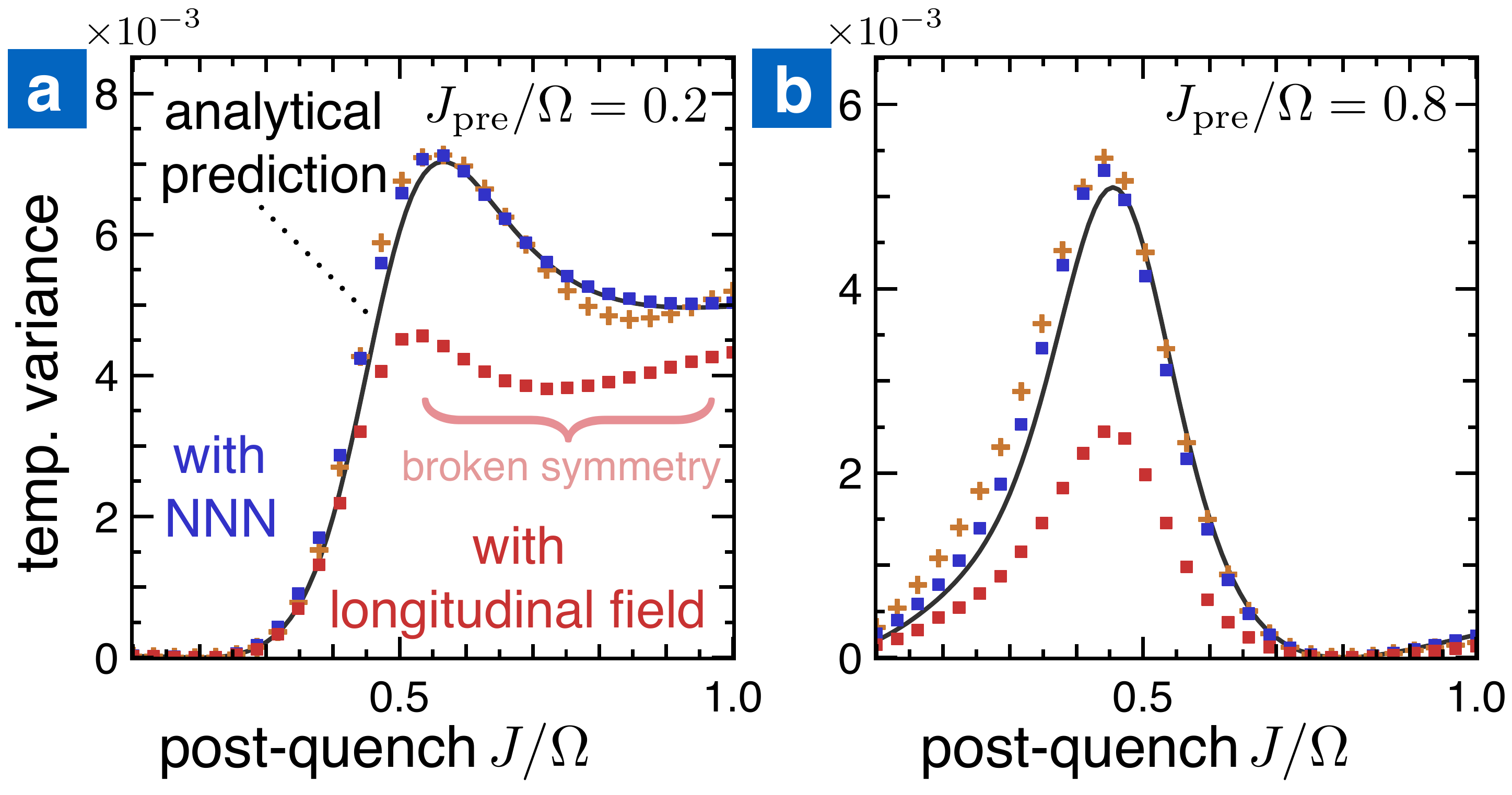}

\protect\caption{\label{fig:2PerturbationsAvaVsNum}Temporal variance $\sigma_{A}^{2}$
for $N=8$ and different pre-quench parameters. The squares show results
from numerical exact diagonalization. Integrability is weakly broken
by $J_{\text{NNN}}/\Omega=0.01$ (blue) or a longitudinal magnetic
field $J_{x}/\Omega=0.01$ (red). The black line shows the analytical
predictions. With the pre- or post-quench parameters in the ferromagnetic
phase and non-zero $J_{x}$ there are stronger deviations due to the
broken $\hat{\sigma}_{x}\protect\mapsto-\hat{\sigma}_{x}$ symmetry.
Decreasing the perturbation further (orange crosses) {[}to $J_{x}/\Omega=2\cdot10^{-4}$
in (a) and $J_{x}/\Omega=5\cdot10^{-4}$ in (b){]} leads to better
agreement. }
\end{figure}

Writing the Ising Hamiltonian in the form of Eq.~(\ref{eq:TransverseIsingHamiltonianSpinLanguage}),
we can give explicit expressions in terms of the ``magnetic field''
directions before ($\vec{b}'_{k}$) and after ($\vec{b}_{k}$) the
quench:
\begin{eqnarray}
{\rm IPR}(k) & = & \frac{1}{2}\left(1+\left(\vec{b}_{k}\vec{b}'_{k}\right)^{2}\right)\\
w(k) & = & \frac{\left(1-b_{zk}^{2}\right)\left(1-\left(\vec{b}_{k}\vec{b}'_{k}\right)^{2}\right)}{16\,{\rm IPR}(k)}
\end{eqnarray}
We find a very good agreement between the analytical expressions derived
here and the numerical results for finite system sizes (Fig.~\ref{fig:2PerturbationsAvaVsNum}).We
can now employ these expressions to discuss how $\kappa$ and $\mathcal{C}$
depend on the quench parameters (Fig.~\ref{fig:DecayConstantAndPrefactor}).
We note especially the non-analytic dependence on the post-quench
parameter at the quantum critical point.

\emph{Other variants} \textendash{} For other observables, similar
calculations can be done. For example, for $\sigma_{x,j}\sigma_{x,j+1}$,
the result is the same up to the change $\hat{S}_{k}^{+}\hat{S}_{k}^{-}\mapsto2\cos(k)\hat{S}_{k}^{+}\hat{S}_{k}^{-}-\sin(k)\hat{S}_{yk}$
in Eq.~(\ref{eq:w(k) equation}). The model discussed here thus affords
an example where the conjectured (and numerically observed) suppression
of fluctuations, exponential in system size, can be studied analytically
in detail.

Other integrability-breaking interactions can be analyzed along the
same lines. For example, consider a weak longitudinal field $\hat{V}_{x}=J_{x}\sum_{j=1}^{N}\sigma_{x,j}$.
Due to the $\hat{\sigma}_{x}\mapsto-\hat{\sigma}_{x}$ symmetry of
the unperturbed system, the first order energy corrections vanish
in the paramagnetic phase. Higher order contributions can still lead
to integrability breaking. In the paramagnetic phase, this yields
a good agreement with our analytical prediction (\ref{fig:2PerturbationsAvaVsNum}).
In the ferromagnetic phase, $\hat{V}_{x}$ breaks the inversion symmetry.
Thus the change in the eigenstates is large and the deviations become
significant (Fig.~\ref{fig:2PerturbationsAvaVsNum}). Still, if we
decrease the perturbation to sufficiently small values (orange crosses)
we once again get a good agreement with the analytical prediction
(see also Suppl. Mat.).

\emph{Conclusions} \textendash{} We have identified a new regime for
quench dynamics of finite-size (``mesoscopic'') weakly non-integrable
many-particle systems, where the fluctuations are suppressed exponentially
in system-size, in contrast to the integrable case. We have presented
a strategy to obtain analytical results for the steady-state long-time
limit.

We expect that the basic mechanism discussed here should apply whenever
one starts from an effectively non-interacting model (where the many-particle
energy can be written as a sum over independent contributions) and
introduces a perturbation that lifts the resulting massive degeneracies.
For finite systems, the perturbation can be weak enough that its effect
on the energies is the main effect, while the many-particle energy
eigenstates are still those of the unperturbed system: even an infinitesimal
interaction is sufficient to lift the degeneracies and will thus lead
to a completely different behaviour of the fluctuations in the long-time
limit (although the time-scale for reaching this limit will of course
diverge as the interaction tends to zero!).

\emph{Acknowledgments} \textendash{} We thank Marcos Rigol, Lea Santos
and Aditi Mitra for discussions. T.K. acknowledges financial support
from the Helmholtz Virtual Institute \textquotedblleft New states
of matter and their excitations''.

\bibliographystyle{apsrev4-1}
\bibliography{references}

\section*{Appendix}

\section{Energy correction}

\begin{figure}
\includegraphics[width=1\columnwidth]{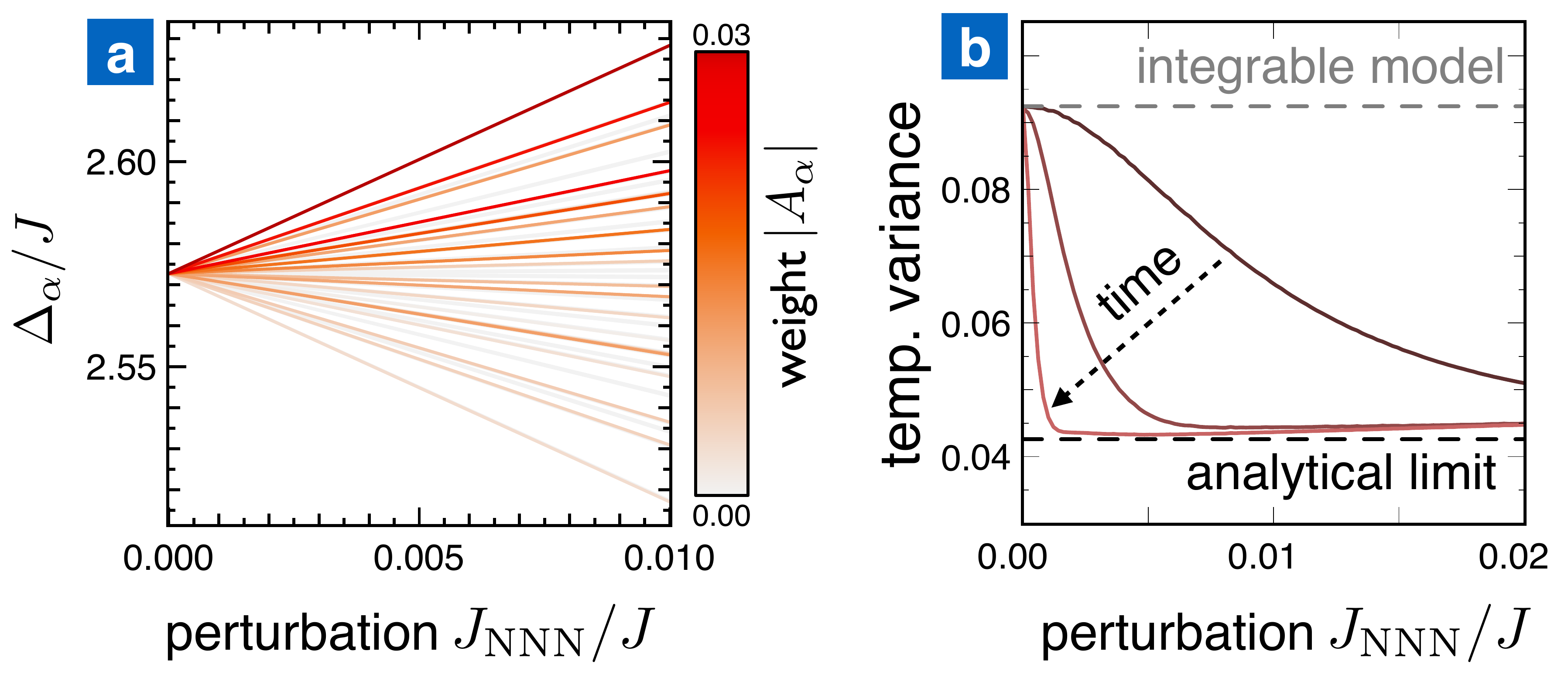}

\protect\caption{\label{fig:GapSplitting-1}Many-particle dephasing. (a) Splitting
of a highly degenerate transition energy $\Delta_{\alpha}$. Here
$\alpha=(f,i)$ is a double index corresponding to the final and initial
state of the transition. The parameters are the same as in figure
\ref{fig:Raw dynamics}, with $N=8$. (b) Temporal variance $N\sigma_{A}^{2}$
as a function of $J_{NNN}$. The three colored curves represent three
different temporal averaging intervals, which start at the time of
the quench and end at $JT=40$, $200$ and $1000$. The longer the
time, the more transition energies are resolved until the analytical
prediction in equation (\ref{eq:ManyBodyDephasing for weak non-integrable models})
is reached.}
\end{figure}

Starting from the fermionic representation of the transverse Ising
Hamiltonian, one can write down the transition energy corrections
to linear order in the coupling:

\begin{equation}
\delta\Delta_{fi}(k)=J_{{\rm NNN}}\sum_{\lambda}\sum_{k'\neq k}F^{(\lambda)}(k,k')W_{k}^{(\lambda)}(n_{k})W_{k'}^{'(\lambda)}(n_{k'})\,.
\end{equation}
The sum over $\lambda$ corresponds to the different terms present
in the fermionic representation of $\hat{V}_{\text{NNN}}$ and for
each $k'$ there is a correction that depends on the respective occupation
numbers. This is very similar to Hartree-Fock corrections for the
interacting Fermi gas. For our purposes, the precise form of $F$
and $W$ is not important, although it can be written down explicitly.
The resulting splitting of a single degenerate transition energy is
plotted in figure \ref{fig:GapSplitting-1}.

\section{Derivation of the analytical many-particle dephasing prediction \label{appendix: MPD derivation - exchanging sum and product}}

In this section we will derive the many-particle dephasing formula
\ref{eq:ManyBodyDephasing for weak non-integrable models} for the
temporal fluctuations of the observable $\hat{A}=\hat{\sigma}_{+,j}\hat{\sigma}_{-,j}$.
First consider the general equation
\begin{equation}
\sigma_{A}^{2}=\sum_{m\neq n}\left|\Psi_{m}^{*}A_{mn}\Psi_{n}\right|^{2}\label{eq:general expression for the variance}
\end{equation}
which was mentioned in the main text. We start by simplifying the
magnitude of the overlap \ref{eq:GeneralOverlapFormula} for two different
configurations $m\neq n$
\begin{eqnarray*}
\left|A_{mn}\right|^{2} & =\frac{1}{N^{2}}\sum_{k,k'>0} & \left\langle \varphi(m,k)\left|\hat{S}_{zk}+1\right|\varphi(n,k)\right\rangle ^{*}I_{k}I_{k'}\\
 &  & \cdot\left\langle \varphi(m,k')\left|\hat{S}_{zk'}+1\right|\varphi(n,k')\right\rangle \\
 & =\frac{1}{N^{2}}\sum_{k>0} & \left|\left\langle \varphi(m,k)\left|\hat{S}_{zk}+1\right|\varphi(n,k)\right\rangle \right|^{2}I_{k}
\end{eqnarray*}

In order to understand this, recall that $I_{k}$ causes all initial
and final spectators to match expect for the wave number $k$. Hence $I_{k}I_{k'}$
can be nonzero only if either the wave numbers are the same or if
$k\neq k'$ and $m=n$. The latter drops out because only different
configuration contribute to the temporal fluctuations.

Inserting into equation \ref{eq:general expression for the variance}
and resolving the Kronecker deltas in $I_{k}$, we obtain
\begin{eqnarray*}
\sigma_{A}^{2} & = & \frac{1}{N^{2}}\sum_{k>0}\sum_{\left\{ \varphi_{k'}=\pm_{k'}\right\} }\left|\left\langle \varphi_{k}\left|\hat{S}_{zk}+1\right|-\varphi_{k}\right\rangle \right|^{2}\\
 &  & \cdot\left|\left\langle \psi_{k}|\varphi_{k}\right\rangle \right|^{2}\left|\left\langle -\varphi_{k}|\psi_{k}\right\rangle \right|^{2}\prod_{\tilde{k}\neq k,\tilde{k}>0}\left|\left\langle \varphi_{\tilde{k}}|\psi_{\tilde{k}}\right\rangle \right|^{4}
\end{eqnarray*}
As in the main text, $|-_{k}\rangle$ corresponds to the ground state
and $|+_{k}\rangle$ to the excited state in the $(k,-k)$ subspace
after the quench. The second sum is over the set of all possible initial
configurations $\left\{ \varphi_{k'}=\pm_{k'}\right\} $. Because
of the Kronecker deltas the final configurations differ only in the
single $k$-sector from the first sum, resulting in $-\varphi_{k}$. 

Next comes the crucial part, exchanging the sum of all configurations
with the product of all positive wave numbers. The result is the product
of all wave numbers $k'$, with each factor containing the sum of
the configurations in the $(k',-k')$-section. Taking special care
of the $k$-sector, we obtain
\[
\begin{aligned}\sigma_{A}^{2}=\frac{1}{N^{2}}\prod_{\tilde{k}>0}\left(\left|\left\langle +_{\tilde{k}}|\psi_{\tilde{k}}\right\rangle \right|^{4}+\left|\left\langle -_{\tilde{k}}|\psi_{\tilde{k}}\right\rangle \right|^{4}\right)\\
\cdot\sum_{k>0}\frac{2\left|\left\langle +_{k}\left|\hat{S}_{zk}+1\right|-_{k}\right\rangle \right|^{2}\left|\left\langle \psi_{k}|+_{k}\right\rangle \right|^{2}\left|\left\langle -_{k}|\psi_{k}\right\rangle \right|^{2}}{\left|\left\langle +_{k}|\psi_{k}\right\rangle \right|^{4}+\left|\left\langle -_{k}|\psi_{k}\right\rangle \right|^{4}}
\end{aligned}
\]

Using the definitions in the main text, we finally arrive at
\[
\sigma_{A}^{2}=\frac{1}{N}\frac{8\sum_{k>0}\omega(k)}{N}\exp\left[\sum_{k>0}\ln\mathrm{IPR}(k)\right]
\]

This corresponds to equation (\ref{eq:ManyBodyDephasing for weak non-integrable models})
in the main text.

\section{Symmetry breaking due to a longitudinal magnetic field\label{appendix: Longitudinal field}}

In the main text, we show results for a longitudinal magnetic field
as a weak perturbation. We found good agreement in the paramagnetic
phase, but deviations in the ferromagnetic phase. These deviations
occur because the longitudinal field breaks the inversion symmetry.
Here we discuss why we still find a good agreement for sufficiently
weak perturbations.

Starting in the ferromagnetic phase and the thermodynamic limit, there
are two degenerate ground states \cite{0022-3719-4-15-024}. One with
all spins pointing along $+x$, and one with all spins along $-x$.
Once the system size $N$ becomes finite, both states can be connected
by $N$ spin flips. The transverse field $\Omega\sigma_{z}$ thus
leads to an exponentially small energy gap that scales as $(\Omega/J)^{N}$.
This gap separates the two new lowest-lying eigenstates, which are
the symmetric and antisymmetric superposition of the two polarized
states. These correspond to the even and odd total particle number
states in the fermionic picture.

In our analytics we assumed the system to be in the even subspace.
A longitudinal magnetic field, stronger than the energy gap caused
by the transverse field, mixes the symmetric and anti-symmetric subspaces.
Thus the eigenstates differ significantly from the unperturbed case.
This leads to deviation that could be corrected in principle, by using
the correct eigenstates in the analytics. However, if the perturbation
is sufficiently small the even and odd subspaces will stay well separated
and the analytical results derived in the main text stay valid.
\end{document}